\documentclass[twocolumn,showpacs,amsmath,amssymb]{revtex4}

\usepackage{dcolumn}% Align table columns on decimal point
\usepackage{bm}% bold math
\usepackage{bbm}
\usepackage{amsfonts}

\def\>{\rangle}
\def\<{\langle}
\def\ket#1{|#1\>}
\def\bra#1{\<#1|}
\def\braket#1#2{\< #1 | #2 \>}

\def\tr{{\,{\rm tr}}}

\def\im{{\,{\rm Im}\,}}

\begin{document}

\title{Estimating purity in terms of correlation functions}
\author{Toma\v z Prosen$^1$, Thomas H. Seligman$^2$ and Marko \v Znidari\v c$^1$}
\affiliation{${}^1$ Physics Department, FMF, University of Ljubljana, Ljubljana, Slovenia\\
${}^2$ Centro de Ciencias F{\'\i}sicas, University of Mexico (UNAM), Cuernavaca,
Mexico}
\date{\today}
\begin{abstract}
We prove a rigorous inequality estimating the purity of a reduced density matrix of a composite quantum
system in terms of cross-correlation of the same state and an arbitrary product state.
Various immediate applications of our result are proposed, in particular concerning Gaussian wave-packet propagation
under classically regular dynamics.
\end{abstract}

\maketitle

The autocorrelation function of time evolution, also known
as {\em survival probability} or as {\em fidelity} in the context of
echo dynamics \cite{fidelity}, is an important tool
in the discussion of quantum systems. 
Indeed the autocorrelation 
function contains a vast amount of information, but analysis in terms of 
the more general concept of 
cross correlation functions can bring additional insight. This has been recently demonstrated  
for spectral statistics \cite{Gorin}. In this note we shall seek a better understanding of 
the evolution of entanglement in terms of cross correlation functions.
This is particularly attractive, because entanglement is at the very root of 
quantum mechanics, and we shall therefore test the usefulness of cross 
correlation functions in this context.
Note that the first experimental test 
on the value of cross correlations for the analysis of
spectral statistics was carried out in a microwave experiment with classical fields \cite{Schaefer}
rather than in the context of quantum mechanics.

Entanglement indicates to what extent the state under consideration can be 
written as a product of states in two subsystems selected due to their interest in the 
physical context. For pure states, entanglement is reflected in the properties of the reduced 
density matrix for any of the two subsystems. While it is tempting to use some form 
of entropy to describe this property, this makes analytic work very 
difficult. We shall therefore use the purity of a subsystem \cite{Zurek} as a 
measure of entanglement. Purity is defined as the trace of the square of 
the reduced density matrix. The fact, that we use an analytic function, allows 
to obtain the basic inequalities with cross correlations, which are the main result of this note.
To explore to what extent these inequalities can be exhausted, we shall 
turn to two examples, which first called our attention to this problem.

First, considering the time evolution under an integrable Jaynes Cummings 
Hamiltonian of a wave packet forming a product state, the group of M.C. Nemes 
\cite{Nemes} observed a strong maximum in the purity after half the period of 
the classical orbit around which 
the packet was constructed \cite{Nemes}. It is intuitively clear that the autocorrelation 
function will show a revival only after a full period, and thus cannot contribute an explanation 
for this phenomenon.
Note that the Jaynes Cummings Hamiltonian
\begin{equation}
H= \hbar \omega a^\dagger a+ \hbar \epsilon J_{\rm z} + \frac{\hbar G}{\sqrt{2J}} (a J_{+} +
a^\dagger J_{-})  
\label{eq:JC}
\end{equation} 
describing an oscillator with standard annihilation operator $a$ and SU(2) spin $\vec{J}$,
has a mirror symmetry given by 
\begin{equation}
a\to -a, \quad J_\pm \to - J_\pm, \quad J_z \to J_z,
\label{eq:sym}
\end{equation}
which is also evident in the figures 1 and 2 of Ref.~\cite{Nemes}. We may then conjecture that 
the oscillations seen in the linear entropy, defined as one minus purity, are due to maxima in the 
autocorrelation functions for full periods and maxima in the cross correlation functions 
with the mirror image of the initial state for the half-period.

Second, following the time evolution of a similar packet under echo 
dynamics with the same Hamiltonian and a slight detuning as perturbation,
we found that coherent states conserved 
purity to higher order in $\hbar$ than other, e.g. random states \cite{Marco}.
Here the conjecture is that the cross correlation function with the 
classically transported image of 
the original packet will yield an explanation. 

These examples serve to illustrate our analytic results in the 
framework of a model, whose use is 
widespread in atomic physics and quantum optics, in particular for Rydberg atoms in 
cavities and for atoms in Paul traps with externally driven fields. The 
first case will become rather obvious once the inequalities are derived, and for the second case we 
shall derive the result within the linear response approximation.

Consider a composite system with a Hilbert space
\begin{equation}
{\cal H}={\cal H}_1\otimes {\cal H}_2,
\end{equation}
consisting of two factor spaces ${\cal H}_{1,2}$ which may have either
finite or infinite dimensions.

Let $\ket{\psi}$ be an arbitrary pure state of a composite system which,
after tracing over the subsystem $2$, defines a reduced density operator
over the subsystem $1$
\begin{equation}
\rho_1 = \tr_2 \ket{\psi}\bra{\psi}.
\end{equation}
We can prove the following general inequality estimating the purity 
\begin{equation}
I[\rho_1] = \tr_1 \rho_1^2
\end{equation}
of a reduced state $\rho_1$.
\\\\
{\bf Theorem:} The following inequality holds
\begin{equation}
|\braket{\phi}{\psi}|^4 \le I[\rho_1].
\label{eq:ineq}
\end{equation}
for any product state $\ket{\phi} = \ket{\phi_1}\otimes\ket{\phi_2}$.
\\\\
{\bf Proof:}  
Uhlmann's theorem \cite{Uhlmann76} 
states for pure states $\rho = \ket{\psi}\bra{\psi}$ and 
$\sigma = \ket{\phi}\bra{\phi}$ that
\begin{equation}
|\braket{\phi}{\psi}|^2 = \tr \rho \sigma \le 
\tr_1[ \tr_2\rho \tr_2\sigma] = \bra{\phi_1}\rho_1\ket{\phi_1}.
\label{eq:uhlmann}
\end{equation}
Then, applying Cauchy-Schwartz inequality for operators
$|{\rm tr}(A^\dagger B)|^2 \le {\rm tr}(AA^\dagger){\rm tr}(BB^\dagger)$ we 
get
\begin{equation}
|\tr_1[\tr_2\rho\tr_2\sigma]|^2 \le \tr_1[\tr_2\rho]^2 \tr_1[\tr_2\sigma]^2
= I[\rho_1].
\label{eq:cs}
\end{equation}
Combining inequalities (\ref{eq:uhlmann}) and (\ref{eq:cs}) we get the
result (\ref{eq:ineq}).
\\\\
{\bf Corollary 1:} An interesting special case of the above result is obtained
if $\ket{\phi}=\ket{\psi(0)}$ is an initial disentangled state of a 
unitary quantum {\em time evolution} $\ket{\psi}=\ket{\psi(t)}$. Then our
results says that the entanglement growth as measured by purity
is bounded from below by the autocorrelation function 
$|\braket{\psi(0)}{\psi(t)}|^4$.
\\\\
{\bf Corollary 2:} A more general situation arises if $\ket{\phi}$ is any factorizable state 
as required by the theorem. Then we obtain that the growth of entanglement as measured by the purity
is bounded from below by all cross correlation functions of the type
$|\braket{\phi}{\psi(t)}|^4$.
\\\\
The second corollary includes the first as a special case and is the general result 
we need in order to apply cross correlation functions to understand oscillations 
or slow growth of the entanglement. 

To explore the possibility to approach equality in the second corollary (and the main theorem)
we have to look at the eigenvalues of the reduced density matrix (which are identical for both subspaces
\cite{x1}). 
As the purity is the trace of an analytic function of the
reduced density matrix the optimal function $\phi$ to establish cross correlation with,
is a product of the eigenfunctions corresponding to the largest eigenvalue.
Note that the eigenvalues are the same for the reduced density matrix in both subspaces. 
If we denote this eigenvalue
by $1-\delta$ the cross correlation function will take the value $|\braket{\phi}{\psi(t)}|^4=(1-\delta)^2$.
Actually, purity $I$ will obey the slightly sharper inequality 
\begin{equation}
(1-\delta)^2+\delta^2\,\ge\, I\,\ge\, (1-\delta)^2+\frac{\delta^2}{\min({\rm dim}({\cal H}_1), 
{\rm dim}({\cal H}_2))-1},
\label{eq:ineq2}
\end{equation}
where ${\rm dim}$ indicates the dimension of the corresponding Hilbert space.
We obtain this inequality by assuming the two extreme cases: either the missing
intensity $\delta^2$ is concentrated in a single product state or it is evenly distributed
among all such states. The former provides the upper bound and the latter the lower bound.
As the lower limit is larger than the value $(1-\delta)^2$ we obtain
from corollary 2 that equality in (\ref{eq:ineq}) can only be reached for $I=1$.
Note that the inequality (\ref{eq:ineq}) provides a fairly sharp lower bound if purity is 
close to 1 because then $1-I \sim \delta$ while $I-|\braket{\phi}{\psi(t)}|^4 \sim \delta^2$,
{\em i.e.} the error of the lower bound is quadratic in the deviation of purity from unity.

We can now hope to find sets of functions $\phi_i$ such that the cross-correlation with these
will explain the behavior of purity as long as the latter is near to $1$. Yet, to be 
illuminating we must be able to construct such a set without diagonalizing the density matrix.
If we have a symmetry group which acts separately on the two subspaces, this set is trivially
generated by applying the symmetry operations to the initial function $\psi(0)$. For the
example discussed in the introduction a cross-correlation with the function obtained 
by applying the reflection (\ref{eq:sym}) to $\psi(0)$, will have maxima at $1/2$, $3/2$, $\ldots$
of the full period. We shall now turn to the other example we mentioned,
namely the slow decay of purity for coherent states in an integrable system. This question was 
discussed in \cite{Marco} for arbitrary Gaussian packets in the context of echo dynamics, and 
illustrated again in the Jaynes Cummings model.

We can follow the line of considering cross correlations a little further in this context if we consider the 
quantity $\tr_1[\tr_2\rho\,\tr_2\sigma]=R(\rho_1,\sigma_1)$ considered in Eq.~(\ref{eq:uhlmann}). For $\rho=\rho(t)$ 
and $\sigma=\rho(0)$ this quantity was introduced as reduced fidelity in \cite{Marco2} implying a reduced 
autocorrelation function if we do not treat an echo situation. 
With this notation we have
$I(\rho_1)\,\ge\,R^2(\rho_1,\sigma_1)\,\ge\, |\braket{\phi}{\psi}|^4$, but in addition the second identity is 
fulfilled if $\phi$ is chosen as the eigenfunction of the largest eigenvalue of $\rho_1$.
By choosing $\rho_1=\rho(t)$ and $\sigma$ arbitrary, we have an additional cross correlation that we may use. 
The basic advantage of this quantity is, that it is, just like purity, defined on the subspace alone,
and may therefore be handier in some situations. As far as approaching the identity with purity is concerned this quantity has no advantages.

We shall next follow the same line of reasoning to find an optimized factorizable function 
$\phi$ to understand the behavior of the purity when it is near identity
for a general Gaussian wave-packet
\begin{equation}
\braket{\vec{x}}{\psi} = 
C\exp(\frac{i}{\hbar}[(\vec{x}-\vec{X})\cdot A (\vec{x}-\vec{X}) + 
\vec{P}\cdot\vec{x}]),
\end{equation}
centered at $(\vec{X},\vec{P})$ and having the shape described by
the (generally complex) matrix 
\begin{equation}
A = 
\begin{pmatrix}
A_{11} & A_{12}\cr
A_{21} & A_{22}\cr 
\end{pmatrix}
\end{equation}
corresponding to a division of a $d=d_1 + d_2$ dimensional
configuration space into $d_1$ and $d_2$ dimensional parts 
$\vec{x}=(x_1,x_2)$. The purity of a reduced wave-packet
$\rho(x_1,x'_1) = \int d x_2
\braket{x_1,x_2}{\psi}\braket{\psi}{x'_1,x_2}$
is $I=\int d x_1 d x'_1 |\rho(x_1,x'_1)|^2$ which can be evaluated
in terms of a $2d$ dimensional Gaussian integral \cite{Marco}
\begin{equation}
I = (\det \im A)
\left|\begin{matrix}
\im A_{11} & \frac{\rm i}{2} A_{12}^*  & 0 & -\frac{\rm i}{2} A_{12} \cr
         \frac{\rm i}{2} A_{21}^*  & \im A_{22} & -\frac{\rm i}{2} A_{21} & 0 \cr
         0 & -\frac{\rm i}{2} A_{12} & \im A_{11} & \frac{\rm i}{2} A^*_{12} \cr
         -\frac{\rm i}{2} A_{21} & 0 & \frac{\rm i}{2} A_{21}^*  & \im A_{22}\cr 
\end{matrix}\right|^{-1/2},
\label{eq:purcs}
\end{equation}
where the vertical lines indicate the determinant of a matrix.
According to our result, this expression should be bounded from below
by cross correlation with any disentangled state which we again choose to 
be a Gaussian packet centered at the same point in phase space
\begin{equation}
\braket{\vec{x}}{\phi} = 
D\exp(\frac{i}{\hbar}[(\vec{x}-\vec{X})\cdot B (\vec{x}-\vec{X}) + 
\vec{P}\cdot\vec{x}]),
\label{eq:packet}
\end{equation}
with
\begin{equation}
B = 
\begin{pmatrix}
B_{11} & 0\cr
0 & B_{22}\cr 
\end{pmatrix}.
\label{eq:Bform}
\end{equation}
Straightforward evaluation of Gaussian integral yields for the
cross correlation
\begin{equation}
|\braket{\phi}{\psi}|^4 = \frac{(\det \im A)(\det \im B)}{|\det (A - B^*)|^2}
\label{eq:cc}
\end{equation}
which, as we have shown, is strictly smaller than (\ref{eq:purcs})
for any $B$ of the form (\ref{eq:Bform}).

One may now ask the following question: 
Which packet (\ref{eq:packet}) optimizes the inequality (\ref{eq:ineq})?
In general this is a complicated optimization problem, however we may
solve it asymptotically when purity is close to $1$. In that case,
off diagonal elements of the shape matrix are small, so we may write
$A_{12}=\epsilon Z_{12}, A_{21} = \epsilon Z_{21}$ where $\epsilon$ is
small. It turns out that the inequality is optimized to leading order
in $\epsilon$ if we take $B_{11}=A_{11}, B_{22}=A_{22}$, namely in that
case $|\braket{\phi}{\psi}|^4=1 - c_1 \epsilon^2 + {\cal O}(\epsilon^4)$,
$I = 1 - c_2 \epsilon^2 + {\cal O}(\epsilon^4)$ where $c_1=c_2$.

We note that the cross-correlation (\ref{eq:cc}) is manifestly $\hbar-$independent.
It may be used to compute dynamical, time dependent cross-correlation function of a propagating
Gaussian packet $\ket{\psi(t)}$ (as long as linearized dynamics is a good approximation), as we know that the 
time dependence of the shape matrix $A=A(t)$ is given by the ratio of two pieces of the classical
monodromy matrix \cite{heller}. 
We have thus shown that to lowest order we can easily optimize the reference wave function 
$\ket{\phi}$ though it will clearly have to vary with time. We could first 
simply use the transported original (and factorizable) state $\ket{\psi(0)}$ centered around the new position,
i.e. $B=A(0)$, where $\braket{\phi}{\psi(t)}$ may be called transported autocorrelation function.
This would not optimize the lower bound, but for short times it might be quite good, and it would
certainly make the bound $\hbar$ independent.
Furthermore, one could apply the bound (\ref{eq:cc}) to the case of echo dynamics where liner response
approximation is valid even for long times if the perturbation is sufficiently small (see \cite{Marco}).

We may summarize our results by stating that we found a general inequality between the fourth power
of cross (and auto) correlation functions and purity. Furthermore, we were able to assert, that the 
lower limit is almost identical to the value, if the purity is near to one. Thus any such value of 
purity can be explained by the fact that the evolving state acquires a large overlap with a 
factorizable state, and the difference between purity and our lower bound is of second order in the deviation 
of purity from identity. The knowledge of this factorizable state may contribute to the understanding
of the slow decay or renewed maxima (also known as recoherence) for the purity.
Particular examples of such situations are given in the case of symmetry group whose representation 
factorizes in the variables in which we wish to split the spaces, and in the case of coherent states.

Financial support by the Ministry of Education, Science and Sports of Slovenia  
and from projects IN-112200, DGAPA UNAM, Mexico,  41000 CONACYT Mexico and
DAAD19-02-1-0086, ARO United States is gratefully acknowledged.


\begin{thebibliography}{99}
\bibitem{fidelity} H.~M.~Pastawski {\em et al.} Phys. Rev. Lett. {\bf 75}, 4310 (1995);
P.~R.~Levstein {\em et al.} J. Chem. Phys. {\bf 108}, 2718 (1998);
R.~A.~Jalabert and H.~M.~Pastawski, Phys.~Rev.~Lett.~ {\bf 86}, 2490 (2001);
T.~Prosen, Phys.~Rev.~E {\bf 65}, 036208 (2002);
T.~Prosen and M.~\v Znidari\v c, J.~Phys.~A {\bf 35}, 1455 (2002);
Ph.~Jacquod {\em et al.} Phys.~Rev.~E {\bf 64}, 055203(R) (2001); 
N.~R.~Cerruti and S.~Tomsovic, Phys.~Rev.~Lett. {\bf 88}, 054103 (2002);
F.~M.~Cucchietti {\em et al.}, Phys.~Rev.~E {\bf 65}, 046209 (2002);
D.~Wisniacki and D.~Cohen, Phys.~Rev.~E {\bf 66}, 046209 (2002).

\bibitem{Gorin} T.~Gorin and T.~H.~Seligman Phys.~Rev.~E {\bf 65}, 026214 (2002).

\bibitem{Schaefer} R.~Sch\" afer {\em et al.} J.~Phys.~A {\bf 36}, 3289 (2003).

\bibitem{Zurek} W.~H.~Zurek, Physics Today {\bf 44}, 36 (1991);
W.~H.~Zurek and J.~P.~Paz, Physica {\bf 83D}, 300 (1995).

\bibitem{Nemes} 
R.~M.~Angelo, K.~Furuya, M.~C.~Nemes, and G.~Q.~Pellegrino, 
%Phys.~Rev.~E {\bf 60}, 5407 (1999); 
Phys.~Rev.~A {\bf 64}, 043801 (2001).

\bibitem{Marco} T.~Prosen, T.~H.~Seligman and M.~\v Znidari\v c, 
{\tt quant-ph/0204043} (Phys. Rev. A {\em in press}).

\bibitem{Uhlmann76} A.~Uhlmann, Rep.~Math.~Phys. {\bf 9}, 273 (1976).

\bibitem{x1} H.~Araki and E.~H.~Lieb, Comm. Math. Phys. {\bf 18}, 160 (1970) 

\bibitem{Marco2}  M.~\v Znidari\v c and T.~Prosen,
J.~Phys.~A {\bf 36}, 2463 (2003)

\bibitem{heller} E.~J.~Heller, in {\em Chaos and Quantum Physics}, ed. A. Voros {\em et al}
(North-Holland, Amsterdam 1991).

\end{thebibliography}
\end{document}